\documentstyle[twocolumn,seceq]{jpsj}
\input epsf.tex

\def\runtitle{Spin, Charge and Quasiparticle Gaps in the One-Dimensional 
Kondo Lattice  with $f^{2}$ Configuration}
\def\runauthor{Shinji {\sc Watanabe}
, Yoshio {\sc Kuramoto}, Tomotoshi
{\sc Nishino} and Naokazu {\sc Shibata}}

\title
{
Spin, Charge and Quasi-Particle Gaps in the One-Dimensional \\
Kondo Lattice  with $f^{2}$ Configuration
}

\author
{
Shinji {\sc Watanabe}\footnote{E-mail: watanabe@cmpt01.phys.tohoku.ac.jp},
Yoshio {\sc Kuramoto}, Tomotoshi {\sc Nishino}$^1$ and
Naokazu {\sc Shibata}$^2$}

\inst
{
Department of Physics, Tohoku University, Sendai 980-8587\\
$^1$Department of Physics, Faculty of Science, Kobe University,
Rokkodai, Kobe 657-8501,\\
$^2$
Institute of Applied Physics, University of Tsukuba, Ibaraki 305-3006,
}

\recdate
{
\today
}

\abst
{The ground state properties of the one-dimensional Kondo lattice
 with an $f^{2}$ configuration at each site
are studied by the density matrix renormalization group method.
At half-filling, competition between the Kondo exchange $J$ and the antiferromagnetic intra f-shell exchange $I$ leads to reduction of energy gaps for spin,  
quasi-particle and charge excitations. 
The attractive force among conduction electrons are induced by the 
competition and the bound state is formed. 
As  $J/I$ increases the $f^{2}$ singlet gives way to the Kondo singlet as the dominant local correlation.
The remarkable change of the quasi-particle gap is driven by the change of 
the spin-1/2 excitation character from the itinerant one to the localized one. 
Possible metal-insulator transition is discussed which may occur as
the ratio $J/I$ is varied by reference to mean-field results in the $f^{2}$ 
lattice system and the two  impurity Kondo system. 
}

\kword
{Crystalline-electric-field singlet, Kondo singlet, 
density matrix renormalization group, Kondo lattice, 
${\ f^{2}}$ Kondo lattice, 
heavy fermion, two impurity system
}

\begin{document}
\sloppy
\maketitle

\section{Introduction}
One of the most important problems concerning heavy fermion systems is to understand how the fermionic quasi-particle states are formed.
In contrast with accumulated information on
Anderson and Kondo lattices with a localized electron per site (called $f^1$ configuration hereafter),  
much less is known about systems with an ${\ f^{2}}$ configuration.
The latter systems, which are called the ${\ f^{2}}$ lattice in this paper, are
relevant to uranium based compounds with $5f^{2}$ configuration ($\rm U^{4+}$) and also to some 
praseodymium compounds with $4f^{2}$ configuration ($\rm Pr^{3+}$).
With even number of localized electrons,
a crystalline electric field (CEF) singlet state can be
realized where the entropy vanishes
at the ground state even though interactions with conduction electrons
or
 with f-electrons at other sites are absent.
This situation is in striking contrast with the case of cerium
compounds with
$4f^{1}$ configuration ($\rm Ce^{3+}$); 
because of the Kramers degeneracy, the entropy would remain at zero
temperature if Ce ion were isolated.
Thus in reality the $4f^{1}$ system chooses either a
magnetically ordered state or the delocalized Fermi liquid state via a
collective Kondo effect.

In an $f^{2}$ lattice,  
the itinerant state is also possible provided that  hybridization
between conduction and f electrons is large enough.
Thus both the localized f-electron picture and band picture can be a
starting point to understand actual compounds with $f^{2}$
configurations.
The most interesting situation occurs when the energy scale of
the CEF singlet state is comparable to that of the itinerant one.
Then the  competition for stability between both states should play a
crucial
role to determine the low energy physics.
We expect that investigation of the competition may provide a key
to understand the mysterious phenomena of real systems.
For example, in $\rm URu_{2}Si_{2}$ the CEF singlet model accounts for
gross
features of highly anisotropic susceptibility and magnetic
transition~\cite{rf:m1}.
The high temperature phase is metallic with the Kondo effect observed
in
the resistivity.
At 17.5 K the specific heat shows a large jump, and
antiferromagnetically ordered moments are observed by neutron
scattering at lower temperature~\cite{rf:15}.
Surprisingly, the magnitude of
the ordered moments is only $0.04\mu_{\rm B}$ which does not reconcile
with the
large jump of the specific heat.
Below 1.2 K, the superconducting phase is
realized in the presence of the tiny magnetic moment.

On  the theoretical side,
  the competition between the CEF singlet and the
Kondo singlet has been studied for the impurity system by the scaling
theory,
in which the reduction of the energy scale due to the competition is
demonstrated.~\cite{rf:13}
In the $f^{2}$ lattice system, the first order phase transition
between the CEF singlet and itinerant states at zero temperature was
derived by the mean field theory~\cite{rf:0}.
Thus it is natural to ask to which extent the latter result depend on
the approximation scheme.
In this paper, we study the competition 
by the the density matrix renormalization group (DMRG) method,
taking the minimal model for the $f^{2}$ lattice system.
The DMRG method offers the most accurate means to study the
ground state numerically.

  The remainder of this paper is organized as follows:
 In $\S$2, we introduce a one-dimensional $f^2$ Kondo lattice model
and
inspect the role of parameters involved.
In $\S$3, we explain how we implement  the density matrix
renormalization group (DMRG) method~\cite{rf:6,rf:20}.
In $\S$4 numerical results are presented and the low energy properties
are discussed.
In $\S$5 we discuss the metal-insulator transition making reference to
mean field results in the $f^{2}$ lattice system and the exact results
for the two impurity Kondo system.
Finally, we summarize the paper in $\S$6.

\section{Model}
\begin{figure}
\begin{center}
\epsfxsize=8cm \epsfbox{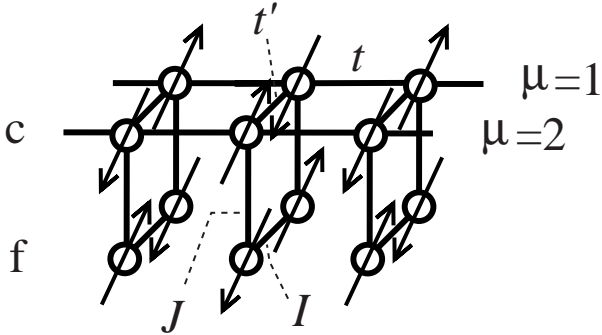}
\end{center}
\caption{Schematic picture of the one-dimensional
$f^{2}$ Kondo lattice model.
The index $\mu \ (=1,2)$ specifies two channels of conduction electrons and
f-electrons.
Conduction electrons are on the upper floor (indicated by "c")
and f-spins are on the lower floor (indicated by "f").
}
\label{fig:ladder1}
\end{figure}

We introduce an $f^{2}$ lattice model in one dimension as follows:
\begin{eqnarray}
\label{eq:hamil}
       H =&-&\sum_{ij\sigma}\sum_{\mu,\nu=1,2}t_{ij}^{\mu\nu}
  (c^{\dagger}_{i\mu\sigma}c_{j\nu\sigma}+{\rm H.c.})
 + J\sum_{i\mu}{\mib S}^{\rm f}_{i\mu} \cdot {\mib S}^{\rm c}_{i\mu}
\nonumber \\
&+&I\sum_{i}{\mib S}^{\rm f}_{i1} \cdot {\mib S}^{\rm f}_{i2},
\end{eqnarray}
where $i$ is the site index, and $\mu$ and $\nu$ denote two channels of
both conduction and f-electrons. In the first term $c^{\dagger}_{i\mu\sigma}
(c_{j\nu\sigma})$ is the creation (annihilation) operator of a conduction
electron with spin $1/2$.
We take the transfer integral in the following form:
\begin{eqnarray}
\label{eq:tt}
t_{ij}^{\mu\nu}=t\delta_{i,j+1}\delta_{\mu,\nu}+t'\delta_{i,j}(1-\delta_{\mu,\nu}),
 \nonumber
\end{eqnarray}
where $t$ is the transfer within each channel, and $t'$ is
the transfer between the two channels at each site (see Fig.\ref{fig:ladder1}).
In eq.(\ref{eq:hamil})
${\mib S}^{\rm f}_{i\mu}$ is the localized f-spin operator with spin
$1/2$, and
${\mib S}^{\rm c}_{i\mu}=(1/2) \sum_{\alpha\beta}c^{\dagger}_{i\mu\alpha}
{\mib \sigma}_{\alpha\beta} c_{i\mu\beta}$
is the spin operator of $\mu$ channel conduction electrons at site
$i$.
The third term with $I\ge0$ causes the singlet-triplet splitting at each site.
This term is introduced to simulate the CEF splitting with the singlet as the ground state.

Here we are interested in the half-filled case where the total number
of
conduction electrons is twice the number $L$ of unit cells:
$$
N\equiv \sum_{\mu=1,2}\sum_{i\sigma}
\langle
c^{\dagger}_{i\mu\sigma}
c_{i\mu\sigma}
\rangle
=2L.
$$
The Hamiltonian is reduced to two sets of the Kondo lattice model
(KLM) if both
the interchannel transfer $t'$ and the  $f^{2}$  coupling $I$
vanish~\cite{rf:18}. 
At half-filling the ground state of the $f^{1}$ KLM is
an insulator where the charge gap is
always larger than the spin gap for any positive $J$.
In the $f^{2}$ lattice model, with $I$ much less than $J$,
the ground state of eq.~(\ref{eq:hamil}) is close to
the direct product of the Kondo singlets on each channel.
Then the ground state is insulating.
On the other hand,
when $I$ is much larger than $J$, the ground state  should tend to
the direct product of the free fermion state and the  $f^{2}$  singlet
states.
The latter ground state is metallic.

\begin{figure}
\begin{center}
\epsfxsize=8cm \epsfbox{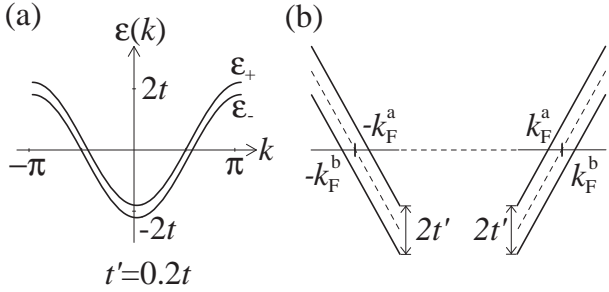}
\end{center}
\caption{(a) Energy dispersion of conduction electrons in the
one-dimensional 
$f^{2}$ Kondo lattice model.
The transfer along chain of each channel is $t$ and
the transfer between the two channels on site is set to be $t'=0.2t$.
(b) Enlargement around the Fermi level. $k_{\rm F}^{\rm b}$
($k_{\rm F}^{\rm a}$) is the Fermi wave number of the bonding
(anti-bonding)
conduction band. }
\label{fig:cos}
\end{figure}

It has been proved that the ground state of eq.~(\ref{eq:hamil}) is
spin singlet and unique\cite{rf:1}. 
The model given by eq.~(\ref{eq:hamil}) has the $\rm SO(4)$ 
symmetry consisting of
two $\rm SU(2)$ symmetries in spin and charge degrees of freedom.
Additionally the model is invariant under the local $\rm U(1)$ gauge
transformation since charge fluctuations of f-electrons are absent.
  The model given by eq.~(\ref{eq:hamil}) has four parameters: $t$, $t'$,
$J$, and $I$.
We take $t=1$ for energy units. The transfer between the two channels
$t'$ let two bands of conduction electrons split as
\begin{eqnarray}
\label{eq:cost}
\varepsilon_{\pm}(k)=-2t \cos(k) \pm t'.
\end{eqnarray}
where the Fermi energy is set to zero with lattice constant unity
(see Fig.\ref{fig:cos}(a)).
For $0 \le t' < 2$ the electronic states are occupied from the bottom
of each band
to the Fermi  energy, and
the system has four Fermi  points.

Let us consider qualitatively the relation between the Fermi wave number of
conduction
electrons and the gap formation.
The ${f^1}$ KLM has a single conduction band, and the difference 
$2k_{\rm F}$ of the Fermi wave numbers 
is commensurate with  the half of the reciprocal wave number 
in the half-filled case.   
In this case the charge gap opens by the infinitesimal perturbation $J$ to gain energy.
In the  ${f^2}$ KLM with $I=0$ and $0< t'< 2$,  each
Fermi wave number $\pm k_{\rm F}^\alpha \ (\alpha = a,b)$ is  not commensurate with the reciprocal lattice.
However, the average of two Fermi wave numbers with the same sign is 
independent of the channel transfer $t'$, since two energy bands are split
symmetrically  with respect to the Fermi energy as given by eq.~(\ref{eq:cost}) 
(See Fig.\ref{fig:cos}(b)).
Namely, the sum $2k_{\rm F}^a+2k_{\rm F}^b$
is $2\pi$ and is equal to the reciprocal lattice unit.
In the frame of perturbation theory with respect to $J$, 
the charge gap in the ${f^2}$ KLM should be smaller than
that of the $f^{1}$ KLM since the gap formation needs higher-order
scattering processes in the two bands in contrast with the the two particle process in the
$f^{1}$ KLM.
  For $t' > 2$ the lower band is fully occupied and the upper band is empty.
Then the system shows the single band feature.
We fix $t'=0.2$ throughout this paper and
 change the  $f^{2}$  coupling $I$ and the Kondo coupling $J$.
In this way we study the competition between  $f^{2}$  singlet and
Kondo singlet states taking account of hybridization between different
channels.

\section{Method of Computation}
We use the density matrix renormalization group method\cite{rf:6,rf:20}.
with the finite-size algorithm.
The open boundary condition is used where
the number of states for each block is kept up to 600.
The maximum system size studied is $L=30$.
Keeping large number of states is necessary especially for the small-$J$ region.

We obtain the lowest energy and the eigenvector 
in the target subspace specified by the 
quantum numbers $S^{\rm z}$ and $N$ of the total Hamiltonian 
by the Lanczos method and the reverse iteration method.  
To reduce the computational time we use the vector obtained by the 
Lanczos method as the initial vector in the reverse iteration method. 
In the Lanczos process the eigenvector $|\Psi\rangle$ is obtained by
requiring that the Rayleigh quotient 
$\langle \Psi |H|\Psi \rangle/\langle \Psi |\Psi
\rangle$  converge to minimum. 
As a result, only one or two iterations are sufficient in the whole
parameter region. 
The eigenvalues  $w_\alpha$ and eigenvectors of the density matrix 
are obtained by the Householder method.  
The truncation error is measured by deviation of 
$P_{m}=\sum_{\alpha=1}^{m}w_\alpha$
from unity. 
In the finite-size DMRG method the energy of the total Hamiltonian becomes lower than that in the previous sweeps 
with the same constitution of left and right blocks.
We judge the convergence of the total energy at the same length
of the left and right blocks since  eigenvector is most improved  
in that case.

\section{Low-Energy Excitations}
\subsection{Spin excitation}
\begin{figure}
\begin{center}
\epsfxsize=8cm \epsfbox{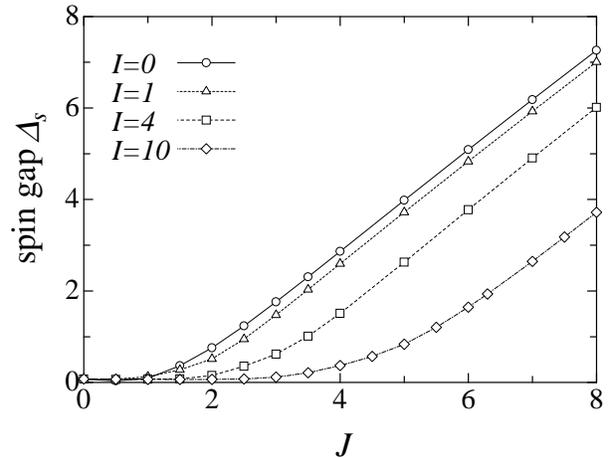}
\end{center}
\caption{Spin gap of the one-dimensional $f^{2}$ Kondo lattice model
with $L=12$.
The transfer $t'$ between the two channels is fixed to $0.2$ in units
of $t$.}
\label{fig:sgp}
\end{figure}

Let us begin with the spin excitation from the ground state.
Figure~\ref{fig:sgp} shows the spin gap against $J$ computed for various values 
of $I$.
The spin gap is obtained from the difference of the ground-state energies
in the subspace of total $S^{z}$ being zero and one with the same total
electron number $2L$; $\Delta_{\rm s}=E(S^{z}=1,N=2L)-E(S^{z}=0,N=2L)$.
The gap is shown for data with $L=12$ since the
truncation error becomes large in the small-$J$ region, and the
reliable value is hard to be obtained for a larger system size.
In this paper, the gap with $L=12$ is shown in the whole parameter
region.
Size dependence of gaps for some parameters is shown later
in Fig.~\ref{fig:gpNs} for information.
We take similar procedure for deriving the charge gap and the
quasiparticle gap.

In the region where $J/I$ is large enough, the spin gap is
proportional to $J$.  This property is common with the
case of the $f^{1}$ KLM\cite{rf:7}.
In the $J \to \infty$ limit, the ground state is a set of local singlet
pairs composed of one conduction electron and the f-electron
at each channel and site.
Then the lowest excitation energy necessary to break up a singlet pair to the triplet one is $J$.
We note that the triplet energy level is split by the off-diagonal matrix element $I/4+I^{2}/8J$ between the channels 1 and 2.
The spin gap is given as the energy difference between
the lower triplet energy and the singlet energy. 
Thus the spin gap becomes smaller owing to the $f^{2}$ coupling $I$.
The second-order perturbation theory with respect to $t$, $t'$ and $I$ gives
the following result:
\begin{eqnarray}
\Delta_{\rm s}=\Delta_{\rm s}^{0}-\frac{I}{4}
-\frac{10t'^{2}}{3J}-\frac{5I^{2}}{32J},
\label{eq:2nd spin gap}
\end{eqnarray}
where $\Delta_{\rm s}^{0}=J-20t^{2}/3J$ is the spin gap of the
$f^{1}$ KLM~\cite{rf:7}.
We remark the relation 
$\Delta_{\rm s} < \Delta_{\rm s}^{0}$ in eq.(\ref{eq:2nd spin gap}).

  Now we turn to the region with small $J$. 
In the $f^{1}$ KLM,
the spin gap behaves as
$\Delta_{\rm s}^{0} \propto \exp (-1/1.4 \rho J)$, where $\rho=1/2\pi
t$ is the
density of states of the conduction band at the Fermi
level\cite{rf:8}.
The spin gap corresponds to the characteristic energy
of the spin component in the Kondo lattice system.
When $I$ is switched on, the moments of f-electrons at each site tend
to form  the local  $f^{2}$  singlet, and the remaining
f-moments are screened by surrounding conduction electrons.
Hence the spin gap stemming from the Kondo effect is reduced by the
competition with the $f^{2}$ singlet.

\subsection{Charge excitation}
\begin{figure}
\begin{center}
\epsfxsize=8cm \epsfbox{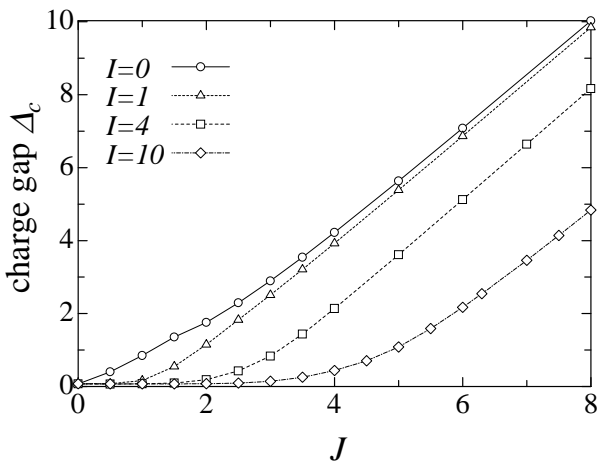}
\end{center}
\caption{Charge gap of the one-dimensional $f^{2}$ Kondo lattice model
with $L=12$ and $t'=0.2$. 
}
\label{fig:cgp}
\end{figure}

Figure~\ref{fig:cgp} shows $J$-dependence of the charge gap for
various $I$.
The charge gap is obtained from the difference of ground-state
energies in the subspace of total number of conduction electrons
being $2L$ and $2L+2$ with the same total spin
$S^{\rm z}=0$; $\Delta_{\rm c}=E(S^{z}=0,N=2L+2)-E(S^{z}=0,N=2L)$.
The hidden $\rm SU(2)$ symmetry in the charge space guarantees that the 
energy difference is the same as the charge gap in the subspace of the total 
electron number fixed to $2L$~\cite{rf:22}. 
The excitation energy to add a conduction electrons to the ground state 
is equivalent to the energy to remove it because of the particle-hole 
symmetry. 
In the region where $J$ is large enough, the charge gap is
proportional to $3J/2$ as in the $f^{1}$ KLM\cite{rf:8}.
In the $J \to \infty$ limit, the energy cost to add one conduction electron
to the ground state is $3J/4$ which is equal to breaking up one local
Kondo singlet.
If another conduction electron is added to the system, the two
electrons added feel short-range repulsive force since the two
electrons cannot transfer if they are on the nearest neighbor sites.
Thus in the bulk limit, the charge gap is proportional to $3J/2$.
The excitation energy $E_{K}(q)$ in the region of large $J$ is given by the second order perturbation theory with respect to $t$,$t'$, and $I$ as
\begin{eqnarray}
\label{eq:Ec}
E_{K}(q) &=& 2 t \cos\left(\frac{K}{2}\right)\cos(q)+
             \frac{2t^{2}}{3J}\cos(K)\cos(2q)
             \nonumber \\
         &-& \frac{3t^{2}}{J}-t'
            -\frac{3t'^{2}}{2J}-\frac{3I^{2}}{8J}+\frac{3t'I}{2J},
\end{eqnarray}
where $K$ is the total momentum and $q$ is the relative momentum.
The charge gap is given by
\begin{eqnarray}
\Delta_{\rm c}=\Delta_{\rm c}^{0}-t'(1-\frac{3I}{2J})
-\frac{t'^{2}}{6J}-\frac{3I^{2}}{16J},
\nonumber
\end{eqnarray}
where $\Delta_{\rm c}^{0}=3J/2-2t+t^{2}/3J$ is the charge gap of the
$f^{1}$ KLM\cite{rf:8}.
It is seen that the charge gap is reduced from the latter system: $\Delta_{\rm c} < \Delta_{\rm c}^{0}$.

In the small-$J$ region the charge gap changes its behavior depending
upon $I$. 
 The charge gap behaves as $J/2$ in the case of $I=0$\cite{rf:8},  
and it is greatly suppressed as $I$ increases. 
This suppression is ascribed to the reduction of the f-moments 
which generate a staggering internal magnetic field on conduction electrons, 
due to the screening by the $f^{2}$ singlet. 
From the data with $I=0,1,4,6,10$ we confirmed that the charge gap is 
always larger than the spin gap in the whole region with $J\ne0$ and
$L=12$. 
However, 
the energy scale of the charge excitation tends to that of the spin excitation as $I$ increases.

\subsection{Single-particle excitation}

\begin{figure}
\begin{center}
\epsfxsize=8cm \epsfbox{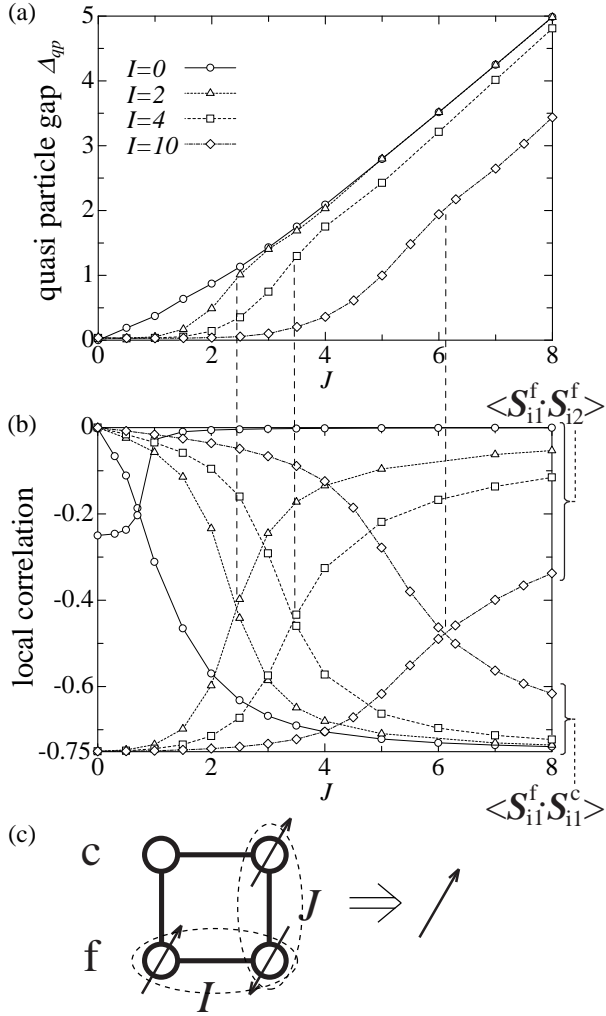}
\end{center}
\caption{(a) Quasi-particle gap of the one-dimensional
$f^{2}$ Kondo lattice model with $L=12$ and $t'=0.2$.
(b) Local correlations at the central site of the system.
Symbols are the same as those in (a). 
(c) Schematic picture of 
spin configurations at a site on which one conduction electron
is added to (or removed from) the ground state.}

\label{fig:qgp}
\end{figure}

Figure~\ref{fig:qgp} (a) shows  the quasi-particle gap against $J$ for
various values of $I$.
The quasi-particle gap is the energy cost to add one conduction
electron to
the ground state:
$\Delta_{\rm qp}=E(S^{z}=1/2,N=2L+1)-E(S^{z}=0,N=2L)$.
In the large-$J$ limit, the quasi-particle gap is proportional to $3J/4$
for any  $I$ as explained above.
From the limit of  large $J$,  the second order perturbation theory 
with respect to $t$, $t'$ and $I$ gives the relation
$\Delta_{\rm qp}=\Delta_{\rm c}/2$.

In the limit of $J \to 0$, the quasi-particle gap is suppressed like the charge gap.
We note that a drastic change appears at a finite value of $J$ when $I$ is
switched on.
Fig.~\ref{fig:qgp} (b) shows $J$-dependence of the local correlations
$\langle \mibs{S}^{\rm f}_{i1} \cdot \mibs{S}^{\rm f}_{i2} \rangle$ and
$\langle \mibs{S}^{\rm f}_{i1} \cdot \mibs{S}^{\rm c}_{i1} \rangle$
at the central site in the system with various values of $I$.
The finite size effect is less than the size of the symbol in
the figure for the whole range of $J$.
The transfer $t'$ between the two channels let f-spins point to opposite
directions on site
even in the $I=0$ case. The crossing point of
$\langle \mibs{S}^{\rm f}_{i1} \cdot \mibs{S}^{\rm f}_{i2} \rangle$ and
$\langle \mibs{S}^{\rm f}_{i1} \cdot \mibs{S}^{\rm c}_{i1} \rangle$
shifts to larger $J$  with increasing $I$.
Comparing Fig.~\ref{fig:qgp} (a) with (b), we find that the point
on which the behavior of the quasi-particle gap changes coincides to the
crossing point of local correlations.

To interpret the change we sketch in Fig.~\ref{fig:qgp} (c) 
the schematic picture of 
spin configurations at a site where one conduction electron is added 
to the ground state.
When $J$ is much less than $I$, the spin $1/2$ has mainly the
character of the conduction electron.
In the opposite case of $J \gg I$, the spin excitation has the
character of the f-electron.
Thus, the character of the spin $1/2$ excitation changes from
the region where the  $f^{2}$  singlet is dominant to the other region where the Kondo singlet is dominant.
Namely, the former is dominantly the itinerant magnetic excitation and the latter the localized magnetic one.
This type of itinerant magnetic excitation is realized in the $f^{2}$ KLM
but not in the $f^{1}$ KLM. 
The quasi-particle gap at the crossing point does not vanish in the
bulk limit.
Note that the crossing point belongs to the insulating phase as will be shown in Fig.~\ref{fig:gpNs}.

\subsection{Attractive force among conduction electrons}
\begin{figure}
\begin{center}
\epsfxsize=8cm \epsfbox{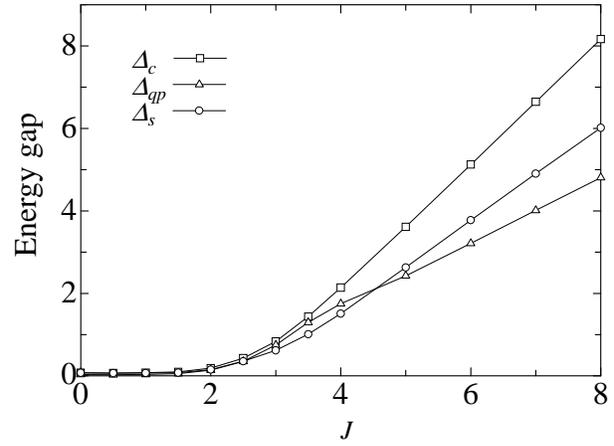}
\end{center}
\caption{Spin gap, quasi-particle gap and charge gap in the one-dimensional
$f^{2}$ Kondo lattice model with $L=12$,  $t'=0.2$ and $I=4$ .
}
\label{fig:3gp}
\end{figure}

Figure~\ref{fig:3gp} shows $J$-dependence of the spin gap $\Delta_{\rm s}$, charge gap $\Delta_{\rm c}$ and the quasi-particle gap $\Delta_{\rm qp}$ with $I=4$. 
The crossing point of local correlations in Fig.~\ref{fig:qgp} (b) is about $J=3.5$. 
We find that the charge gap is less than
twice the quasi-particle gap:
\begin{eqnarray}
\label{eq:2qpltc}
\Delta_{\rm c} < 2 \Delta_{\rm qp}. \nonumber
\end{eqnarray}
This is in marked contrast with the $I=0$ case 
where the charge gap is twice the quasi-particle gap
in the whole-$J$ region just as in the $f^{1}$ KLM\cite{rf:8}.
The difference is due to appearance of the bound state for the charge excitation in the $f^{2}$ KLM.     
Namely attractive force works among quasi-particles, which can be
 understood most simply from the large-$J$ limit as follows: 
If an electron is removed from the ground state, a local Kondo singlet is broken.  Then two remote holes cost the energy $3J/2$.  If they are on the same site, the energy cost is $3J/2-3I/4$ since the $f^2$ singlet can be formed in the latter case.
The difference $-3I/4$ can be regarded as an attractive force among two holes.  
Similar force works for two electrons added to the ground state because of 
the particle-hole symmetry.  
Note that the $-3I/4$ term does not appear in eq.~(\ref{eq:Ec}) since the perturbation theory with respect to $t$, $t'$ and $I$ does not probe the bound state.  
Hence the contribution from the attractive force is only of the order  $L^{-1}$, and is omitted  in eq.~(\ref{eq:Ec}).


It is seen in Fig.~\ref{fig:3gp} that $\Delta_{\rm c}/\Delta_{\rm qp}$ becomes smaller as $J$ decreases at least down to $J=3.5$.
For $J<3.5$, the charge gap is close to the quasi-particle gap and is a little larger 
than the spin gap. 
We confirmed the closeness also in the case of $I$=1,4,6,10 
in the region where 
the $f^{2}$ singlet becomes dominant.

\section{Metal-Insulator Transition}
  Toward understanding the electronic state in the $f^{2}$ lattice system, a necessary step is to clarify the difference from the $f^{2}$
impurity
 system. This impurity system is further related to the two-impurity Kondo
system with $f^1$ configuration at each site.
Namely, the $f^{2}$ impurity system is
the short-distance limit of the two-impurity Kondo system.
Fortunately we have detailed knowledge about the two-impurity system
by the
mean-field theory\cite{rf:2}, the Quantum Monte Carlo\cite{rf:3}, and
numerical renormalization group\cite{rf:4,rf:5}.
In the two-impurity Kondo model,  the change of the ground state
between  the $f^{2}$  singlet and Kondo singlet states occurs as a
quantum phase transition at a finite value of $J/I$.
On the other hand, the change is continuous in the two-impurity
Anderson model with the charge fluctuation of
f-electrons.
Since eq.~(\ref{eq:hamil}) is considered as the extension to the
lattice system
of the two-impurity Kondo model, it is interesting to see whether
the change
between the $f^{2}$ and Kondo singlet states occurs as a phase transition.
In the limit of the  $f^{2}$  singlet state,  spin and charge gaps are
zero, while in the Kondo singlet state both gaps are finite.

Recently, the following has been proved  for one-dimensional systems 
which have the translational invariance and conserve the total electron 
number, parity and time reversal:
The spin (charge) gap is absent when $n_{\uparrow}-n_{\downarrow}$ 
($n_{\uparrow}+n_{\downarrow}$) is not an integer, 
where $n_{\uparrow}$ ($n_{\downarrow}$) is the number of up (down) spin 
electron per unit cell~\cite{rf:17}. 
In the present half-filled case, each number is an integer and
excitation gaps can be either present or absent. 
It is known that at half-filling the Kondo-Hubbard model, where the
conduction electrons have an on-site attraction, has a quasi long-range
order without either spin or charge gaps~\cite{rf:14}.
In that phase, a doubly occupied site and an empty site with respect
to conduction electrons appear by turns.
As a result the spin configuration including f-spin at each site
becomes equivalent to  that in the Heisenberg chain with spin 1/2.
Hence there is no spin gap.
On the other hand the $f^{1}$ KLM has gaps in both spin  and the charge excitations in the half-filled case~\cite{rf:16}.

In the frame of mean-field theory\cite{rf:0},
a first-order phase transition  between the CEF singlet
state
and the Kondo singlet state has been demonstrated at finite $J/I$ for the same model
as given by eq.~(\ref{eq:hamil}).
The mean-field result is independent of the dimensionality of the
system  as long as
the density of states of conduction electrons is smooth 
at the Fermi energy.
This is because only conduction electrons in  the vicinity of the
Fermi level participate dominantly  in the formation of the Kondo
singlet in the region of small $J/I$.
We should, however, note that the first-order transition may be linked to 
the following important feature of the 
mean-filed approximation:
the charge fluctuations of f-electrons 
enter even if the 
average occupation of f electrons is two at each site.
The charge fluctuations is necessary to make finite the following mean
fields:
$\langle f^{\dagger}_{i\mu\sigma}c_{i\mu\sigma} \rangle$ and
$\langle f^{\dagger}_{1\mu\sigma}f_{2\mu\sigma} \rangle$.
On the other hand, in exact theories such as DMRG, these quantities
are zero since the charge fluctuations of f-electrons are completely
suppressed and the conjugate phase fluctuations are fully enhanced.

The effective Hamiltonian with the mean-fields is reduced to the $U=0$
periodic Anderson model with $f^{2}$ configurations.
The ground state of the Hamiltonian is metallic in the limit of $J/I
\to 0$  and
is insulating in the opposite limit of $J/I \to \infty$.
In the intermediate region, a new state
appears in eq.~(\ref{eq:hamil}) when the f-level splitting is larger
than the Kondo temperature.
This state is metallic because the density of states is finite in some region between bonding and anti-bonding f-levels.
In the two impurity Anderson model\cite{rf:3,rf:5}, the corresponding state is responsible for smooth connection of the  $f^{2}$  singlet state to the
Kondo singlet state.
In the $f^{2}$ lattice model, however, this intermediate metallic state is unstable.
Hence the first-order phase transition occurs as $J/I$ increases
in the mean-field theory\cite{rf:0}.
In the exact theory without charge fluctuation of f electrons, there is no f-level splitting and there is no intermediate state.
Namely, the Anderson-type model to solve finally in the mean-field theory may display  different physics from that in  the original Kondo lattice model.

\begin{figure}
\begin{center}
\epsfxsize=8cm \epsfbox{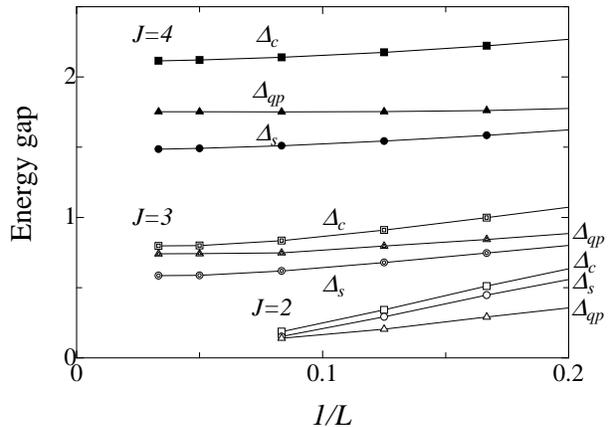}
\end{center}
\caption{Size dependence of spin gap, quasi-particle gap and charge
gap in the one-dimensional 
$\rm f^{2}$ Kondo lattice model ($L=6,8,12,20$ and $30$ for $J=3$ and
$J=4$.
$L=6,8$ and $12$ for $J=2$).
The parameters are $t'=0.2$ and $I=4$.}
\label{fig:gpNs}
\end{figure}

To determine in the DMRG the critical ratio of $J/I$ on which
possible change between the metal and the insulator
occurs, we kept the number of states up to 1500 in the calculation and
used parallelized algorithm to accelerate the convergence.
The size dependence of the gaps for $I=4$ is shown in Fig.~\ref{fig:gpNs}.
It is clear that the size dependence becomes large for small $J$.  In the case of $J=2$, for example, there is no tendency to saturation at $L=12$. 
Unfortunately, however, as $J$ becomes smaller the truncation error
becomes large since the Kondo cloud extends much
and the fluctuations due to the competition with the  $f^{2}$  singlet grow in the case of $I\ne0$.
Because of this computational difficulty, we have no definite answer so far whether the gap collapses at finite $J$ as $J/I$ decreases, or the gap extends to $J=0$.

\section{Conclusions and Discussions}
We have investigated the ground state property and low energy
excitations
in the $f^{2}$ Kondo lattice model and have shown the followings:

  First, the spin gap, charge gap and quasi-particle gap are reduced
by the competition between the local  $f^{2}$  singlet and the Kondo
singlet.
This result implies the characteristic energy scales of spin and
charge
components are decreased from the value of the $f^{1}$ KLM. 
In the whole parameter region the charge gap is larger than the spin
gap and
in the small-$J/I$ region the charge gap is reduced so as to be close
to
the spin gap.

  Secondly, the quasi-particle gap changes its behavior between the
region
where
the  $f^{2}$  singlet is dominant and the region where the Kondo
singlet is dominant.
This is because the character of the spin $1/2$ excitation changes
between these regions; from the itinerant character to the localized
one.

Thirdly, we find a mechanism which makes two conduction electrons added to the ground state feel the attractive force due to competition between the exchange interactions $I$ and $J$.
As a result, the bound state is formed in the charge excitation. 

In relation to the superconducting phase, it is interesting how the
attractive force affects conduction electrons off the half-filling.
In the metallic state, two conduction electrons come into the same site
if the  $f^{2}$  coupling $I$ is more effective than the channel
transfer $t'$.
However, 
there is no bare attractive interaction along the chain direction 
between conduction electrons in eq. (\ref{eq:hamil}).
Hence the situation is different from the doped Kondo chain with the Heisenberg coupling of f-spins, 
where  the spin gap phase and the phase separation appear as a result of
attractive intersite interaction among conduction electrons due to the Heisenberg coupling~\cite
{rf:19}.

Since actual $f^{2}$ compounds are mostly  three dimensional,
we finally discuss relevance of the competition between the Kondo singlet and
the CEF singlet
in higher dimensions.
In the ${f^1}$ KLM, the antiferromagnetic order is expected to occur
in the small-$J$ region in the half-filled case for systems with
dimensions larger than two ~\cite{rf:18,rf:12}.
In the ${f^2}$ KLM, both the Kondo
singlet and the  $f^{2}$  singlet tend to screen the magnetic moment
of f-electrons.
Thus the condensation energy of the magnetic order is small even in
higher
dimensions.
Competition between the magnetic order,  the Kondo effect and  the $f^{2}$ singlet formation may then stabilize tiny magnetic moments.

\section*{Acknowledgments}
We are grateful to M. Sigrist and O. Sakai for valuable discussions.
S. W. would like to thank S. Hikihara for useful advice
about DMRG techniques.
A part of the numerical calculation was performed by VPP500 at the
Supercomputer Center of the ISSP, Univ.~of Tokyo.
Parallelized FORTRAN code was executed by SX4R at the Supercomputer
Center, Tohoku University.
This work is supported by CREST from the Japan Science and Technology Cooperations, and by a Grant-in-Aid for Scientific Research from the Ministry of Education, Science and Culture.


\begin{thebibliography}{99}
\bibitem{rf:m1} G. J. Niewenhuys: Phys. Rev. B {\bf 16} (1987) 5260.
\bibitem{rf:15} C. Broholm, H. Lin, P. T. Matthews, T. E. Mason,
W. J. L. Buyers, M. F. Colins, A. A. Menovsky, J. A. Mydosh, J. K. Kjems
: Phys. Rev. B {\bf 43} (1991) 12809.
\bibitem{rf:13} Y. Kuramoto:
{\it Transport and Thermal Properties of f-Electron Systems}, ed.
G. Oomi, H. Fujii and T. Fujita (Plenum Press, New York, 1993)
p. 237.
\bibitem{rf:0} S. Watanabe and Y. Kuramoto: Z. Phys. B {\bf 104} (1997) 535.
\bibitem{rf:6} S. R. White: Phys. Rev. Lett. {\bf 69} (1992) 2863.
\bibitem{rf:20} S. R. White: Phys. Rev. B {\bf 48} (1993) 10345.
\bibitem{rf:18} H. Tsunetsugu, M. Sigrist and K. Ueda: Rev. Mod. Phys.
{\bf69} (1997) 809.
\bibitem{rf:1} H. Tsunetsugu: Phys. Rev. B {\bf55} (1997) 3042.
\bibitem{rf:7} H. Tsunetsugu: Phys. Rev. B {\bf46} (1992) 3175.
\bibitem{rf:22} T. Nishino and K. Ueda: Phys. Rev. B {\bf 47} (1993) 12451. 
\bibitem{rf:8} N. Shibata, T. Nishino, K. Ueda and C. Ishii:
Phys. Rev. B {\bf53} (1996) 8828.
\bibitem{rf:2} B. A. Jones, B. G. Kotliar and A. J. Millis:
Phys. Rev. B {\bf 39} (1989) 3415.
\bibitem{rf:3} R. M. Fye and J. E. Hirsch: Phys. Rev. B {\bf 40} (1989) 4780.
\bibitem{rf:4} B. A. Jones, C. M. Varma and J. W. Wilkins:
Phys. Rev. Lett. {\bf 61} (1988) 125.
\bibitem{rf:5} O. Sakai, Y. Shimizu and T. Kasuya:
J. Phys. Soc. Jpn. {\bf 59} (1990) 2414.
\bibitem{rf:17} M. Yamanaka, M. Oshikawa and I. Affleck:
Phys. Rev. Lett. {\bf79} (1997) 1110.
\bibitem{rf:16} A. M. Tsvelik:
Phys. Rev. Lett. {\bf72} (1994) 1048.
\bibitem{rf:14} N. Shibata, M. Sigrist, E. Heeb:
Phys. Rev. B {\bf56} (1997) 11084.
\bibitem{rf:19} A. E. Sikkema, I. Affleck, and S. R. White:
Phys. Rev. Lett. {\bf79} (1997) 929.
\bibitem{rf:12} M. Vekic, J. W. Cannon, D. J. Scalapino, R. T.
Scaletter,and R. L. Suger:
Phys. Rev. Lett. {\bf74} (1995) 2367.
\end{thebibliography}
\end{document}